\documentclass[12pt,tightenlines,nofootinbib,amsmath,amssymb,aps,prd,notitlepage]{revtex4-1}

\usepackage{graphicx}
\usepackage[caption=false]{subfig}

\usepackage{amsmath,amssymb}
\usepackage{amsfonts,amssymb,mathrsfs}

\usepackage[bookmarks=false,pdfstartview=FitH]{hyperref}
\usepackage[all]{hypcap}

\usepackage[left=1.1in,right=1.1in]{geometry}

\def\be{\begin{equation}}
\def\ee{\end{equation}}
\def\nn{\nonumber}
\def\f{\frac}
\def\tf{\tfrac}
\def\pl{{\rm Pl}}
\def\lp{\ell_\pl}
\def\b{\bar}
\def\d{\dot}

\def\t{\tilde}
\def\v{\vec}

\def\de{\delta}
\def\ep{\epsilon}
\def\ga{\gamma}

\def\mH{\mathcal{H}}
\def\mO{\mathcal{O}}

\def\mC{\mathcal{C}}

\def\oe{\mathring{e}}

\def\lo{\ell_o}

\usepackage{color}

\begin{document}

\title{The loop quantum cosmology bounce as a Kasner transition}

\author{Edward Wilson-Ewing} \email{edward.wilson-ewing@unb.ca}
\affiliation{Department of Mathematics and Statistics, University of New Brunswick, Fredericton, NB, Canada E3B 5A3}

\begin{abstract}

For the Bianchi type I space-time (vacuum or with a massless scalar field), the loop quantum cosmology bounce can be viewed as a rapid transition between two classical solutions, with a simple transformation rule relating the Kasner exponents of the two epochs.  This transformation rule can be extended to other Bianchi space-times under the assumption that during the loop quantum cosmology bounce the contribution of the spatial curvature to the Hamiltonian constraint is negligible compared to the kinetic terms.  For the vacuum Bianchi type IX space-time there are transformation rules for how each of the parameters characterizing the Kasner epochs change during the bounce.  This provides a quantum gravity extension to the Mixmaster dynamics of general relativity, and may have interesting implications for the Belinski-Khalatnikov-Lifshitz conjecture.

\end{abstract}

\maketitle

\section{Introduction}
\label{s.intro}

One of the main objectives for any theory of quantum gravity is to determine the fate of the singularities predicted by general relativity.  There has been significant progress in this direction for loop quantum cosmology (LQC), where the background independent quantization procedure developed for loop quantum gravity is applied to symmetry-reduced homogeneous Friedmann-Lema\^itre-Robertson-Walker (FLRW) universes \cite{Bojowald:2001xe, Ashtekar:2003hd, Ashtekar:2006wn} and Bianchi space-times \cite{Bojowald:2003md, Chiou:2007sp, MartinBenito:2008wx, Ashtekar:2009vc, MartinBenito:2009qu, Ashtekar:2009um, WilsonEwing:2010rh} (for a review of homogeneous LQC, see e.g.~\cite{Ashtekar:2011ni, Banerjee:2011qu}).  In LQC, the big-bang and big-crunch singularities of the FLRW space-time are resolved by quantum gravity effects and are generically replaced by a non-singular bounce \cite{Ashtekar:2006wn, Ashtekar:2007em, MartinBenito:2009aj, MenaMarugan:2011me, Pawlowski:2014fba}.

In addition, the full quantum dynamics of the expectation values of sharply-peaked states are well approximated by the LQC effective dynamics that include the key LQC effects \cite{Ashtekar:2006wn, Taveras:2008ke, Rovelli:2013zaa}.  For the spatially flat FLRW space-time, the effective Friedmann equation is
\be
H^2 = \f{8 \pi G}{3} \rho \left( 1 - \f{\rho}{\rho_c} \right),
\ee
where $H$ is the Hubble rate, $\rho$ is the energy density of the matter fields, and $\rho_c \sim \rho_{\rm Pl}$ is the critical energy density; the continuity equation is unchanged by LQC.  Clearly, the LQC bounce happens when $H=0$ and $\rho=\rho_c$, and LQC effects are negligible for $\rho \ll \rho_c$.  An important point here is that the quantum gravity corrections predicted by LQC originate in the geometrical sector of the theory; by using some identities it is possible rewrite the effective dynamics as done above with the corrections to the classical Friedmann equation now appearing in the matter sector.

While the LQC dynamics of the Bianchi space-times have not yet been studied in as much detail as the FLRW space-times, preliminary numerical studies indicate that sharply-peaked Gaussian states also undergo a non-singular bounce in Bianchi I space-times \cite{Pawlowski-talk}.  In addition, for solutions of the LQC effective equations for the Bianchi space-times, not only is the singularity replaced by a bounce \cite{Gupt:2012vi, Corichi:2012hy, Corichi:2015ala}, but also geometric invariants like the expansion and shear are bounded \cite{Corichi:2009pp, Singh:2013ava} and all strong singularities are resolved \cite{Singh:2011gp, Saini:2017ipg, Saini:2017ggt}.

Beyond spatially homogeneous space-times, the linearly polarized Gowdy space-time has been studied in a hybrid quantization where the phase space is split into homogeneous and inhomogeneous sectors which are respectively quantized by LQC and Fock techniques \cite{Garay:2010sk, MartinBenito:2010bh, MartinBenito:2010up, deBlas:2017goa}.  In this system also the singularity is resolved insofar as the zero-volume `singular' state decouples from non-singular states under the action of the Hamiltonian constraint operator, and the expansion and shear are bounded in the effective dynamics \cite{Tarrio:2013ija}.

It may be possible to gain further insight into the fate of singularities in inhomogeneous space-times through the Belinski-Khalatnikov-Lifshitz (BKL) conjecture that in the approach to a space-like singularity, spatial derivatives become negligible compared to time-like derivatives, and contributions from matter fields to the dynamics also become negligible (with the exception of a massless or kinetic-dominated scalar field) \cite{Belinski:1969, Belinski:1970ew, Andersson:2000cv, Uggla:2003fp}.  (For a reformulation of the BKL conjecture in terms of variables suited to loop quantum gravity, see \cite{Ashtekar:2008jb, Ashtekar:2011ck}.)  If the conjecture is correct, then near a space-like singularity the equations of motion at each point become those of a vacuum Bianchi space-time.

This conjecture has spurred detailed studies of the dynamics of the vacuum Bianchi space-times in classical general relativity which found that in the approach to the cosmological singularity the Bianchi space-time undergoes a sequence of oscillations between near-Kasner solutions (with the Kasner solution being the vacuum Bianchi I space-time) \cite{Belinski:1969, Belinski:1970ew, Misner:1969hg, Misner:1969ae}.  Indeed, the spatial curvature can often be neglected (at which times the spatially flat Bianchi I solution is a good approximation) and only becomes important for short periods of time when it causes a transition to rapidly occur between two different near-Kasner solutions.  As a result, the dynamics of Bianchi space-times near the cosmological singularity can be approximated by a sequence of Kasner solutions (or by a sequence of Bianchi I solutions with a massless scalar field in the presence of a kinetic-dominated scalar field \cite{Belinski:1973zz, Berger:1999tp}).

The BKL conjecture then suggests that in the approach to a generic space-like singularity, neighbouring points decouple and the solution at each point is given, to a good approximation, by a sequence of Kasner solutions.  While numerical studies provide evidence for the BKL conjecture in a number of settings (and also give some idea of the relative time-scales between the duration of the near-Kasner solutions and the rapid transitions between them) \cite{Weaver:1997bv, Berger:1998xx, Berger:1998vxa, Garfinkle:2003bb} (see \cite{Berger:2014tev} for a recent review), the validity of the BKL conjecture in general remains an open question; however, some limitations are known: not all singularities in general relativity are space-like \cite{Dafermos:2017dbw}, and even in the approach to a space-like singularity there are typically some isolated `recurrent spike surfaces' where space-like derivatives remain comparable to time-like derivatives and cannot be neglected \cite{Rendall:2001nx, Andersson:2004wp, Lim:2007ta, Lim:2009dg, Heinzle:2012um, Coley:2016yuk}.  Nonetheless, numerical studies indicate that the BKL conjecture may hold at generic points near space-like singularities.

If the BKL conjecture holds at generic points in the presence of quantum gravity effects, then it will be possible to extend the singularity resolution results obtained for the Bianchi space-times to the inhomogeneous case.  In addition, the LQC dynamics for Bianchi space-times would also provide important insights into how quantum gravity effects may modify the dynamics of inhomogeneous space-times.  This second point is the motivation for this paper, which aims to provide a clearer understanding of the LQC dynamics for Bianchi space-times, and thereby of the quantum gravity corrections predicted by LQC for space-times where the BKL conjecture holds.

Numerical solutions of the LQC effective dynamics for the Bianchi I space-time show that during the LQC non-singular bounce from a contracting to an expanding space-time, a classical solution is a very good approximation away from the immediate vicinity of the bounce point (with one solution before, and a different solution after).  During the bounce itself, the Kasner exponents (which parametrize the pre-bounce classical solution) do change, but very rapidly reach new constant values for the post-bounce classical solution within $t_{\rm Pl}$, up to a factor of order unity \cite{Gupt:2012vi}.  This suggests that the LQC bounce in Bianchi space-times, at least according to the LQC effective dynamics, can be viewed as a very rapid transition between different classical Bianchi I solutions.  In fact, it turns out that the transition rule relating the Kasner exponents of the Bianchi I solutions before and after the bounce has a very simple form.

In classical general relativity, the dynamics for contracting Bianchi space-times can be approximated by a sequence of Bianchi I solutions linked by the BKL transition map \cite{Belinski:1969, Belinski:1970ew}; in LQC, when the space-time curvature nears the Planck scale there will be a new type of Kasner transition due to the LQC bounce, with a different transition map.  After the bounce, LQC effects will rapidly become negligible and then the dynamics of the now expanding space-time can once more be approximated by a sequence of Bianchi I solutions linked by the usual BKL transition map.

The dynamics for the Bianchi space-times in general relativity will be reviewed in Sec.~\ref{s.class}, in order to set the notation and provide a self-contained summary of the results that will be extended in Sec.~\ref{s.lqc} to LQC.  A brief discussion of potential implications for the BKL conjecture is given in Sec.~\ref{s.bkl}.

\section{Classical Theory}
\label{s.class}

Bianchi space-times have a four-dimensional manifold $\Sigma \times \mathbb{R}$, where the metric on the spatial manifold $\Sigma$ is homogeneous with a three-dimensional symmetry group that is simply transitive, parametrized by three Killing vector fields $\mathring\xi^a_i$.  The fiducial triads $\oe^a_i$, which satisfy the relations $\oe^a_i \oe_b^i = \de^a_b$ and $\oe^a_i \oe_a^j = \de_i^j$, commute with the Killing vector fields, $[\oe_i, \mathring\xi_j]^a = 0$, and satisfy the algebra
\be
[\oe_i, \oe_j]^a = C^k{}_{ij} \, \oe^a_k,
\ee
where the structure constants $C^k{}_{ij}$ depend on the specific Bianchi space-time.  For example, for Bianchi I $C^k{}_{ij}=0$, for Bianchi II $C^1{}_{23} = -C^1{}_{32} = 1$ and all other $C^k{}_{ij} = 0$ and for Bianchi IX $C^k{}_{ij}=\ep^k{}_{ij}$.  For more on Bianchi space-times, see \cite{Ryan-Shepley, Jantzen:2001me}.

In this paper, I will consider the diagonal Bianchi I, Bianchi II and Bianchi IX space-times, for which the spatial metric is $q_{ab} = e_a^i e_{bi}$ and the physical co-triads are
\be \label{def-e}
e_a^i = a_i(t) \, \oe_a^i, \qquad {\rm no~sum~over~} i,
\ee
the $a_i(t)$ are the directional scale factors which without loss of generality can be taken to be positive.

In the Hamiltonian formulation of general relativity, a convenient set of conjugate variables are the densitized triads $E^a_i = \sqrt q \, e^a_i$ and the extrinsic curvature $K_a^i = K_{ab} e^{bi}$.  (It would also be possible to use Ashtekar-Barbero variables, but in this case the calculations are simpler in terms of $E^a_i$ and $K_a^i$.)  For the diagonal Bianchi I, Bianchi II and Bianchi IX space-times, the densitized triads and extrinsic curvature can be parametrized by
\be \label{def-EK}
E^a_i = \f{p_i}{\lo^2} \sqrt{\mathring{q}} \,\, \oe^a_i, \qquad K_a^i = \f{K_i}{\lo} \, \oe_a^i, \qquad {\rm no~sum~over~} i,
\ee
where $p_i = a_j a_k \lo^2$ (with $j \neq k$ both different from $i$) and $K_i$ depend only on time, while $\lo$ is the cubic root of the coordinate volume of the spatial manifold%
\footnote{For the Bianchi IX space-time with the structure constants $C^k{}_{ij} = \epsilon^k{}_{ij}$ the coordinate volume of the spatial manifold is $\lo^3 = 16 \pi^2$ (see, e.g., \cite{WilsonEwing:2010rh}).  On the other hand, there is no preferred value for $\lo$ in the Bianchi I and Bianchi II space-times and for these space-times it is possible to choose $\ell_1, \ell_2, \ell_3$ to be different \cite{Ashtekar:2009vc, Ashtekar:2009um}, although this is not necessary here.  Note also that if the spatial manifold is non-compact, then it is necessary to restrict integrals to a fiducial cell, with coordinate volume $\lo^3$, but the dynamics of the space-time are independent of the choice of the fiducial cell.},
$\lo^3 = \int \! \sqrt{\mathring{q}}$.  (Classically, $K_i = \lo \, d a_i/dt$, with $t$ being the proper time, but this relation is a result of the dynamics of general relativity and does not necessarily hold in effective LQC.)

The Poisson brackets $\{K_a^i(\v x), E^b_j(\v y)\} = 8 \pi G \, \de_a^b \, \de^i_j \, \de^{(3)}(\v x - \v y)$ of the full theory induce on the reduced phase space the Poisson bracket
\be
\{K_i, p_j\} = 8 \pi G \, \de_{ij}.
\ee
Given this parametrization for $(E^a_i, K_a^i)$, the constraints encoding spatial diffeomorphisms and rotations of the triads are automatically satisfied.  The remaining scalar constraint,
\be
\mH = - \, \f{1}{16 \pi G} \left( 2 \f{E^a_i E^b_j}{\sqrt q} K_{[a}^i K_{b]}^j + \sqrt q \,\, {}^{(3)}\!R \right) + \f{1}{2 \, \sqrt q} \Big( \pi_\phi^2 + E^a_i E^{bi} \partial_a \phi \partial_b \phi \Big) \approx 0,
\ee
(allowing for a minimally coupled massless scalar field $\phi$, with momentum $\pi_\phi = p_\phi \sqrt{\mathring{q}} / \lo^3 = \sqrt{q} \, \partial_t \phi$) reduces for a homogeneous space-time with the parametrization \eqref{def-EK} to
\be
\mH = - \, \f{\sqrt{\mathring{q}}}{8 \pi G \lo^3 \sqrt{p_1 p_2 p_3}} \Big[p_1 p_2 K_1 K_2 + p_1 p_3 K_1 K_3 + p_2 p_3 K_2 K_3 \Big] 
+ \f{\sqrt{\mathring{q}} \, U \! (p_i)}{\sqrt{p_1 p_2 p_3}} + \f{\sqrt{\mathring{q}} \, p_\phi^2}{2 \lo^3 \sqrt{p_1 p_2 p_3}},
\ee
where the potential $U(p_i) = - p_1 p_2 p_3 \, {}^{(3)}\!R / 16 \pi G \lo^3$ depends on the Bianchi model.  The Poisson bracket for the homogeneous scalar field $\phi(t)$ is $\{\phi, p_\phi\} = 1$.

In this paper, I will only consider the massless scalar field as a possible matter content (which includes the vacuum case of $p_\phi=0$) since this is the only type of matter field that grows at the same rate as anisotropies in a contracting universe, and so is the only matter field typically expected to contribute to the dynamics of the space-time in the high-curvature regime.  (An exception to this is an ekpyrotic field with an equation of state greater than 1, which will dominate over anisotropies, but in this case the space-time will be driven to the FLRW solution \cite{Cai:2013vm}, and then there is no need to include anistropies.)  Nonetheless, it is possible to include other types of matter fields, for example the effect of a dust field is studied in \cite{Ali:2017qwa}.

Dynamics are generated by the Hamiltonian constraint $\mC_H = \int N \mH$: for any observable $\mO$, $d \mO / d\tau = \{\mO, \mC_H\}$, with $\tau$ determined by the choice of the lapse $N$.  For Bianchi space-times, the Hamiltonian constraint operator has a particularly simple form for $N = \sqrt{p_1 p_2 p_3} / \lo^3$.  It is also convenient to change variables to
\be
\alpha_i = \ln a_i, \qquad \Pi_i = p_j K_j + p_k K_k, \qquad {\rm no~sum~over~} i,j,k {\rm ~and~} i,j,k {\rm ~all~different,}
\ee
for example, $\alpha_1 = \ln a_1 = \tf{1}{2} [\ln (p_2 / \lo^2) + \ln (p_3 / \lo^2) - \ln (p_1 / \lo^2)]$ and $\Pi_1 = p_2 K_2 + p_3 K_3$.  Note that this change of variables is valid for $p_i \neq 0$, i.e., away from the big-bang singularity.  These variables are canonically conjugate, with the Poisson brackets
\be
\{ \alpha_i, \Pi_j\} = -8 \pi G \, \de_{ij}, \qquad \{\alpha_i, \alpha_j\} = \{\Pi_i, \Pi_j\} = 0.
\ee
In these variables and for the lapse $N = \sqrt{p_1 p_2 p_3}/\lo^3 = \exp(\sum_i \alpha_i)$, the Hamiltonian constraint is
\be
\mC_H = \!\! \int \!\! N \mH = - \, \f{1}{16 \pi G \lo^3} \Big[ \Pi_1 \Pi_2 + \Pi_1 \Pi_3 + \Pi_2 \Pi_3 - \f{1}{2} (\Pi_1^2 + \Pi_2^2 + \Pi_3^2) \Big] + U(\alpha_i) + \f{p_\phi^2}{2 \lo^3},
\ee
where due to homogeneity the integral is trivial and gives a factor of $\lo^3$.  Again, the potential $U(\alpha_i)$ depends on the Bianchi space-time.

\subsection{Bianchi I}
\label{ss.kasner}

For the Bianchi I space-time, the spatial curvature ${}^{(3)}\!R$ vanishes, so $U(\alpha_i) = 0$ and
\be \label{ham-kasner}
\mC_H = - \, \f{1}{16 \pi G \lo^3} \Big[ \Pi_1 \Pi_2 + \Pi_1 \Pi_3 + \Pi_2 \Pi_3 - \f{1}{2} (\Pi_1^2 + \Pi_2^2 + \Pi_3^2) \Big] + \f{p_\phi^2}{2 \lo^3},
\ee
(note that both $\Pi_i$ and $p_\phi$ scale as $\lo^3$, so $\mC_H \sim \lo^3$ also).  As a result, for the Bianchi I space-time all of the $\Pi_i$ as well as $p_\phi$ are constants of the motion.  Without any loss of generality, I will assume that $p_\phi \ge 0$ in what follows.  The equations of motion for $\alpha_i(\tau)$ and $\phi(\tau)$ give
\be \label{eom-alpha}
\f{d \alpha_i}{d\tau} = \f{1}{2 \lo^3} (\Pi_j + \Pi_k - \Pi_i), \qquad \Rightarrow \qquad
\alpha_i = \f{(\Pi_j + \Pi_k - \Pi_i) \tau}{2 \lo^3} + \alpha_i^{(0)},
\ee
\be
\f{d\phi}{d\tau} = \f{p_\phi}{\lo^3}, \qquad \Rightarrow \qquad \phi = \f{p_\phi \tau}{\lo^3} + \phi_o,
\ee
where in \eqref{eom-alpha} it is understood that the indices $j \neq k$ are both different from $i$.

This is the complete solution for the dynamics of the Bianchi I space-time in general relativity.  It is possible to make contact with the metric in the Kasner form
\be
ds^2 = -dt^2 + (t - t_o)^{2k_1} (\oe^1)^2 + (t - t_o)^{2k_2} (\oe^2)^2 + (t - t_o)^{2k_3} (\oe^3)^2,
\ee
by changing the time variable from $\tau$ to the proper time $t$, related by $e^{\sum_i\alpha_i} d\tau = dt$,
\be \label{time}
e^{\sum_i (\tf{1}{2 \lo^3}\Pi_i \tau + \alpha_i^{(0)})} = \f{1}{2 \lo^3} \sum_i \Pi_i \, (t-t_o), \qquad
\tau = \f{2 \lo^3}{\sum_j \Pi_j} \left( \ln \f{\sum_i \Pi_i}{2 \lo^3} (t-t_o) - \sum_i \alpha_i^{(0)} \right),
\ee
with $t - t_o > 0$ if $\sum_i \Pi_i > 0$ while $t - t_o < 0$ if $\sum_i \Pi_i < 0$.  Then,
\be \label{def-ai}
a_i(t) = e^{\alpha_i(t)} = A_i \, (t - t_o)^{k_i},
\ee
where the $A_i = (\tf{1}{2 \lo^3} \sum_j \Pi_j)^{k_i} \cdot \exp[\alpha_i^{(0)} - k_i \sum_j \alpha_j^{(0)}]$ are constants, and the Kasner exponents $k_i$ are given by
\be \label{def-ki}
k_i = \f{\Pi_j + \Pi_k - \Pi_i}{\Pi_1+\Pi_2+\Pi_3},
\ee
with $j, k$ different from each other and both different from $i$, e.g., $k_1 = (\Pi_2 + \Pi_3 - \Pi_1)/(\Pi_1 + \Pi_2 + \Pi_3)$.

The mean scale factor is
\be \label{mean-a}
a = (a_1 a_2 a_3)^{1/3} = \left( \f{\sum_i \Pi_i}{2 \lo^3} \, (t - t_o) \right)^{1/3} = \exp \left[ \f{1}{3} \sum_i \left( \f{\Pi_i \, \tau}{2 \lo^3} + \alpha_i^{(0)} \right) \right];
\ee
if $\sum_i \Pi_i > 0$ then the space-time expands as $\tau$ increases and $a \to 0$ as $\tau \to -\infty$, while if $\sum_i \Pi_i < 0$ then the space-time contracts as $\tau$ increases and $a \to 0$ as $\tau \to +\infty$.

The equations of motion for $\phi$ can also be rewritten in terms of $t$,
\be
\phi = k_\phi \, \ln \left( \f{\sum_i \Pi_i (t - t_o)}{2 \lo^3} \right) + \t \phi_o, \qquad k_\phi = \f{2 \, p_\phi}{\Pi_1+\Pi_2+\Pi_3},
\ee
note that the sign of $\sum_i \Pi_i$ determines the sign of $k_\phi$ (given the earlier assumption that $p_\phi \ge 0$).

From the relation in \eqref{def-ki}, it is obvious that the first Kasner relation holds,
\be
\sum_i k_i = 1,
\ee
and the second Kasner relation
\be
\sum_i k_i^2 = 1 - 8 \pi G \, k_\phi^2,
\ee
follows from the Hamiltonian constraint $\mC_H=0$.  Note that in the $k_\phi = 0$ vacuum case, exactly one of the Kasner exponents $k_i$ must be negative.  On the other hand, if $k_\phi^2 \neq 0$, then it is possible for all $k_i$ to be positive, with this becoming more likely as $|k_\phi|$ increases.  In particular, for $|k_\phi| > 1/\sqrt{16 \pi G}$ all $k_i$ must be positive.

\subsection{Bianchi II}
\label{ss.b2}

For the Bianchi II space-time, the potential is an exponential wall in the $+\alpha_1$ direction,
\be \label{U-b2}
U_{II}(\alpha_i) = \lo^3 \: \f{e^{4 \alpha_1}}{32 \pi G}.
\ee
At late times in an expanding Bianchi II space-time, the scale factor $a_1 \to 0$, so $\alpha_1 \to -\infty$ and the potential \eqref{U-b2} becomes negligible.  In this late time and large volume limit, since the potential $U_{II}(\alpha_i)$ is negligible the solution to the equations of motion is the Kasner solution reviewed in Sec.~\ref{ss.kasner}, with the condition $k_1 < 0$ since $a_1 \to 0$ at late times.

Starting at late times with the Kasner solution $(k_i, k_\phi)$, with $k_1 < 0$, and evolving back in time, the potential grows and will eventually become important; at this point the Kasner solution is no longer a good approximation to the full Bianchi II solution.  However, as studied in detail by Belinski, Khalatnikov, Lifshitz and Misner \cite{Belinski:1969, Belinski:1970ew, Belinski:1973zz, Misner:1969hg, Misner:1969ae}, the potential is only important for a short period of time and soon afterwards the Bianchi II solution can again be approximated by a Kasner solution with different parameters $(\t k_i, \t k_\phi)$.  In general relativity, this Kasner solution remains a good approximation to the Bianchi II solution for the remainder of its evolution, which ends at the big-bang singularity when the volume $V = a_1 a_2 a_3 \, \lo^3 = e^{\sum_i \alpha_i} \lo^3 \to 0$.

The relation between the Kasner parameters $(k_i, k_\phi)$ and $(\t k_i, \t k_\phi)$ before and after this Kasner transition can be derived from an explicit solution for the Bianchi II metric, but there is a simpler calculation that also gives the transition rules.  From the form of the potential, it is clear that $\Pi_2, \Pi_3$ and $p_\phi$ are all constants of the motion at all times, including when the potential is important, and it is only $\Pi_1$ that has non-trivial evolution during the transition when $U_{II}(\alpha_i)$ becomes large.  Of course, in either Kasner epoch far from this transition point, $\Pi_1$ is (to a very good approximation) constant.  Therefore, during the Kasner transition, $\Pi_1$ is shifted by some finite amount $\Delta\Pi_1$:
\be
\Pi_1 \to \t \Pi_1 = \Pi_1 + \Delta\Pi_1.
\ee
Clearly, the $(\Pi_i, p_\phi)$ at late times must satisfy the Bianchi I Hamiltonian constraint \eqref{ham-kasner} (since the potential is negligible at these times), and the $(\t \Pi_i, \t p_\phi)$ must also satisfy this same constraint, again since the potential becomes negligible at early times well before the Kasner transition.  The result of requiring that both $(\Pi_i, p_\phi)$ and $(\t \Pi_i, \t p_\phi)$ satisfy the Hamiltonian constraint for the Bianchi I space-time \eqref{ham-kasner} is that
\be
\Delta\Pi_1 \left[ \f{\Delta\Pi_1}{4} - \f{\Pi_2 + \Pi_3 - \Pi_1}{2} \right] = 0,
\ee
which only has two solutions, $\Delta\Pi_1 = 0$ (corresponding to the late-time solution), and
\be \label{pi-map}
\Delta\Pi_1 = 2 (\Pi_2 + \Pi_3 - \Pi_1), \qquad \Rightarrow \qquad
\t \Pi_1 = -\Pi_1 + 2 \Pi_2 + 2 \Pi_3,
\ee
corresponding to the early-time solution on the other side of the Kasner transition.

From the definition of $k_i$ and $k_\phi$, it is easy to obtain the well-known transition rule for the Kasner exponents $k_i$
\be \label{bkl-map}
\t k_1 = \f{-k_1}{1+2 k_1}, \qquad
\t k_2 = \f{k_2 + 2 k_1}{1 + 2 k_1}, \qquad
\t k_3 = \f{k_3 + 2 k_1}{1 + 2 k_1},
\ee
and also for $k_\phi$,
\be \label{kphi-map}
\t k_\phi = \f{\Pi_1 + \Pi_2 + \Pi_3}{3 \Pi_2 + 3 \Pi_3 - \Pi_1} k_\phi = \f{k_\phi}{1 + 2 k_1}.
\ee
Note that since $k_1 < 0$, the denominator is less than 1 and so it follows that $\t k_\phi > k_\phi$ (because $k_\phi$ must be positive in a Kasner solution that is contracting as $\tau \to -\infty$).

Finally, since the potential $U_{II}(\alpha_i)$ for the Bianchi II space-time has a single wall, the system `bounces off' the potential wall only once, and there is only one transition in the Kasner exponents throughout the entire Bianchi II evolution.

\subsection{Bianchi IX}
\label{ss.b9}

The potential for the Bianchi IX space-time is
\be \label{U-b9}
U_{IX}(\alpha_i) = \lo^3 \: \left[ \f{e^{4 \alpha_1} + e^{4 \alpha_2} + e^{4 \alpha_3}}{32 \pi G} - \f{e^{2 (\alpha_1 + \alpha_2)} + e^{2 (\alpha_1 + \alpha_3)} + e^{2 (\alpha_2 + \alpha_3)}}{16 \pi G} \right],
\ee
which will clearly generate richer dynamics than the Bianchi II potential.  Importantly, the potential only becomes important when any one of the $\alpha_i$ nears zero (or equivalently when any of the scale factors $a_i$ nears $1$).  When all $\alpha_i \ll -1$ (i.e., when all $a_i \ll 1$) the potential is negligible and the space-time will be (approximately) Kasner.  Note also that the potential is symmetric under any permutation of $(\alpha_1, \alpha_2, \alpha_3)$.

Assuming that no two of the scale factors have a similar amplitude, then the largest term in $U_{IX}(\alpha_i)$ at any given time will be one of the $e^{4 \alpha_i}/4$ exponential walls (since the $e^{2(\alpha_i+\alpha_j)}$ terms are the geometric mean of two of the $e^{4 \alpha_i}/4$ walls, and will therefore be much smaller unless two of the $\alpha_i$ are the same order of magnitude), and this is precisely the potential for the Bianchi II space-time \eqref{U-b2}.  As a result, Kasner transitions corresponding to permutations of \eqref{bkl-map} will occur whenever one of these terms becomes sufficiently large to affect the dynamics.

The main difference with the Bianchi II space-time is that while only one such Kasner transition occurs during the entire Bianchi II dynamics, many Kasner transitions will occur in a Bianchi IX space-time because, in a contracting space-time, so long as any one of the Kasner exponents is negative the corresponding scale factor will grow until the potential becomes sufficiently large to cause a Kasner transition.  So, in the Bianchi IX space-time, the system can `bounce off' the walls of the potential many times throughout its evolution, each `bounce' corresponding to a Kasner transition given by a permutation of \eqref{bkl-map}.

In the vacuum case, since there must always be one $k_i < 0$, there will be an infinite number of these Kasner transitions as the singularity is approached.  On the other hand, in the presence of a massless scalar field there will be a finite number of Kasner transitions \cite{Belinski:1973zz, Berger:1999tp}.  This is because if $k_\phi \neq 0$ then it is possible for all $k_i$ to be positive, in which case all scale factors will decrease as the singularity is approached and the potential \eqref{U-b9} will remain negligible for the entire remainder of the evolution, so no more Kasner transitions will occur.  Note that for $k_\phi \neq 0$ it is guaranteed that at some point all $k_i$ will be positive, since this is required for $|k_\phi| > 1 / \sqrt{16 \pi G}$, and $|k_\phi|$ increases at each Kasner transition, as can be seen from \eqref{kphi-map}, so eventually $|k_\phi|$ will be sufficiently large so that all $k_i$ must be positive.  (Of course, in some cases the condition that all $k_i > 0$ will be satisfied and Kasner transitions will stop before $|k_\phi|$ reaches the critical value of $1/\sqrt{16 \pi G}$.)

For the vacuum Bianchi IX space-time, there exists a simple parametrization for each approximate Kasner solution given by $(u, p_\Omega, v, \kappa)$, with these parameters defined by \cite{Berger:1996gk} (see also \cite{Chernoff:1983zz} for a related but slightly different choice of parameters)
\be
u = \f{k_{\rm max}}{k_{\rm mid}}, \qquad p_\Omega = \f{1}{6 \lo^3} \sum_i \Pi_i,
\ee
\be
v = \f{k_{\rm mid}}{k_{\rm max}} \cdot \f{k_{\rm min} \alpha_{\rm max} - k_{\rm max} \alpha_{\rm min}}{k_{\rm min} \alpha_{\rm mid} - k_{\rm mid} \alpha_{\rm min}} + 1, \qquad
\kappa = k_{\rm max} \left( \f{\alpha_{\rm min}}{k_{\rm min}} - \f{\alpha_{\rm mid}}{k_{\rm mid}} \right).
\ee
These are all constants of the motion for any Kasner solution (as can be checked from the dynamics of the $\alpha_i$), and therefore the parameters $(u, p_\Omega, v, \kappa)$ will all be (approximately) constant during the periods that the Bianchi IX space-time is well approximated by a Kasner solution, and will only change during the transitions between successive approximate Kasner solutions.  Here $-\tf{1}{3} < k_{\rm min} < 0 < k_{\rm mid} < \tf{2}{3} < k_{\rm max} < 1$ are the Kasner exponents ordered from smallest to greatest and $\alpha_{\rm min}, \alpha_{\rm mid}, \alpha_{\rm max}$ are their respective logarithmic scale factors.  To see that $v$ and $\kappa$ are constant for a Kasner solution, it is helpful to rewrite \eqref{eom-alpha} as, e.g.,
\be \label{alpha-min}
\alpha_{\rm min}(\tau) = 3 \, k_{\rm min} \, p_\Omega \, \tau + \alpha_{\rm min}^{(0)},
\ee
and similarly for $\alpha_{\rm mid}(\tau)$ and $\alpha_{\rm max}(\tau)$.  Then a short calculation shows not only that (for Kasner solutions) all of $(u, p_\Omega, v, \kappa)$ are constant, but also that $u, v > 1$ and $\kappa > 0$ (for $\alpha_i \le 0$).

All three Kasner exponents are determined by $u$,
\be \label{k-from-u}
k_{\rm min} = \f{-u}{1 + u + u^2}, \qquad
k_{\rm mid} = \f{1 + u}{1 + u + u^2}, \qquad
k_{\rm max} = \f{u + u^2}{1 + u + u^2},
\ee
and note that it is possible to rewrite $v$ as
\be
v = - \, \f{1}{\kappa} \Big( \alpha_{\rm max} + (1+u) \alpha_{\rm min} \Big) + 1.
\ee

At each Kasner transition, it is $\Pi_{\rm min}$ that changes following the map \eqref{pi-map} while $\Pi_{\rm mid}$ and $\Pi_{\rm max}$ do not change, although the labels $min, mid, max$ will be interchanged after the transition.  From these considerations, a short calculation shows that after each Kasner transition
\be \label{u-map}
u \to \tilde u =
\begin{cases}
u-1 & {\rm if~} ~ u - 1 > 1, \\
(u-1)^{-1}  & {\rm otherwise},
\end{cases}
\qquad
p_\Omega \to \tilde p_\Omega = p_\Omega \, \f{u^2 - u + 1}{u^2 + u + 1}.
\ee
Note that after the Kasner transition the $min$ and $mid$ labels are exchanged, unless $u - 1 < 1$ in which case the permutation $mid \to min, max \to mid, min \to max$ of the labels is required.  Each period of constant $u$ is known as a Kasner epoch, and a Kasner era refers to the sequence of all consecutive epochs where $u$ decreases by 1; an era change occurs when $u \to \tilde u = 1/(u-1)$.

Furthermore, making the approximation that the Kasner transition occurs instantaneously when the largest $\alpha_i = 0$, a transformation map for $v$ and $\kappa$ can also be derived,
\be \label{v-map}
v \to \tilde v =
\begin{cases}
v+1 & {\rm if~} ~ u - 1 > 1, \\
1 + 1/v & {\rm otherwise},
\end{cases}
\qquad
\kappa \to \tilde \kappa =
\begin{cases}
\kappa & {\rm if~} ~ u - 1 > 1, \\
v \, \kappa/(u-1) & {\rm otherwise}.
\end{cases}
\ee

This map (which provides a good approximation to the dynamics of the Bianchi IX space-time) has been shown to be chaotic \cite{Barrow:1981sx, Chernoff:1983zz}, and it is the portion of the map that sends $u \to \tilde u = 1/(u-1)$ which will drive apart solutions that were initially close.  Also, note that a more precise transition map than \eqref{u-map} and \eqref{v-map} has been found \cite{Berger:1996gk}.  Nonetheless, the qualitative dynamics of an infinite sequence of Kasner oscillations that are highly sensitive to the initial conditions remains the same.

Further, numerical solutions of the full dynamics of the Bianchi IX space-time have been shown to be chaotic by coordinate-independent fractal methods \cite{Cornish:1996yg, Cornish:1996hx}.  (Note that many common measures of chaotic behaviour such as the Lyapunov exponent are coordinate-dependent in that they depend explicitly on the choice of a time variable.  This is why for diffeomorphism-invariant theories like general relativity it is important to use coordinate-independent methods to quantify chaotic behaviour.)

\section{Effective Loop Quantum Cosmology}
\label{s.lqc}

Numerical studies of the effective LQC dynamics of the Bianchi I space-time show that while quantum gravity effects in LQC are sufficiently strong to resolve the big-bang singularity and replace it by a non-singular bounce, these quantum gravity effects are nonetheless very short-lived \cite{Gupt:2012vi}.  In particular, the Kasner exponents are to a very good approximation constant before the bounce, then change suddenly at the bounce point and settle to new constant values very rapidly.  This suggests that it may be possible to view the LQC bounce as a new type of Kasner transition, with a new transition rule different from the BKL map.  This section derives this new transition rule for the LQC bounce for the Bianchi I space-time, which can then be extended to the Bianchi II and Bianchi IX space-times for the case that the potential $U(\alpha_i)$ is negligible during the bounce.

\subsection{Bianchi I}
\label{ss.lqc-kasner}

The LQC effective dynamics for the Bianchi I space-time based on the quantum theory studied in \cite{Ashtekar:2009vc} (this loop quantization of the Bianchi I space-time is preferred due to its consistency for both compact and non-compact spaces, and because the quantum gravity corrections only arise when the space-time curvature is near the Planck scale \cite{Corichi:2009pp}) for $N = V / \lo^3 = \sqrt{p_1 p_2 p_3} / \lo^3 = \exp[\sum_i \alpha_i]$ are generated by the LQC effective Hamiltonian constraint \cite{Ashtekar:2009vc} (see also Appendix C in \cite{Chiou:2007sp})
\begin{align} \label{ch-b1}
\mC_H & = -\f{V^2}{8 \pi G \ga^2 \Delta \lo^3}
\sum_{i \neq j} \sin \b\mu_i \ga K_i \sin \b\mu_j \ga K_j
+ \f{p_\phi^2}{2 \lo^3} \nn \\ &
= -\f{V^2}{8 \pi G \ga^2 \Delta \lo^3}
\Big[ \sin \f{\ga \sqrt\Delta}{2 V} (\Pi_2 + \Pi_3 - \Pi_1)
\sin \f{\ga \sqrt\Delta}{2 V} (\Pi_1 + \Pi_3 - \Pi_2) \nn \\ & \qquad \qquad
+ \sin  \f{\ga \sqrt\Delta}{2 V} (\Pi_2 + \Pi_3 - \Pi_1)
\sin  \f{\ga \sqrt\Delta}{2 V} (\Pi_1 + \Pi_2 - \Pi_3) \nn \\ & \qquad \qquad
+ \sin \f{\ga \sqrt\Delta}{2 V} (\Pi_1 + \Pi_3 - \Pi_2)
\sin \f{\ga \sqrt\Delta}{2 V} (\Pi_1 + \Pi_2 - \Pi_3) \Big]
+ \f{p_\phi^2}{2 \lo^3},
\end{align}
where $\gamma$ is the dimensionless Barbero-Immirzi parameter, $\Delta = 4 \sqrt 3 \, \pi \gamma \lp^2$ is the smallest non-zero eigenvalue of the area operator in LQG, and $\b\mu_i = \sqrt{p_i \Delta / p_j p_k} = \sqrt\Delta \, p_i / V$.

A key point is that $\mC_H$ depends only on the $\alpha_i$ through the combination $V = \exp(\sum_i \alpha_i) \lo^3$, and as a result the equations of motion for all of the $\Pi_i$ will be identical since $\{\Pi_i, V\} = 8 \pi G V$:
\be \label{lqc-pi}
\f{d \Pi_i}{d \tau} = 8 \pi G \, V \, \f{d \, \mC_H}{d V}.
\ee
The explicit form of $d \, \mC_H / dV$ is not important here, what matters is that the equations of motion for all of the $\Pi_i$ are identical.

As a result, given a classical solution $(\Pi_i, p_\phi)$ that approximates the LQC solution before the bounce, and a different classical solution $(\t\Pi_i, \t p_\phi)$ that approximates the LQC solution after the bounce, the parameters of these solutions will be related by $\t p_\phi = p_\phi$ and $\t\Pi_i = \Pi_i + \Delta\Pi$, where all the $\t\Pi_i$ are shifted by the same amount given by the integral of the right side of \eqref{lqc-pi}.  However, it is not necessary to evaluate this integral.

Instead, it is enough to notice that the classical solutions before and after the bounce must both satisfy the classical Hamiltonian constraint \eqref{ham-kasner}, and this requires that the shift $\Delta\Pi$ satisfy
\be
\Delta\Pi \left( 2 \sum_i \Pi_i + 3 \Delta\Pi \right) = 0,
\ee
which again has only two solutions.  $\Delta\Pi = 0$ correponds to the pre-bounce solution, and the post-bounce solution is given by
\be \label{pi-lqc-map}
\Delta\Pi = - \f{2}{3} \sum_i \Pi_i, \qquad \Rightarrow \quad \t \Pi_i = \Pi_i - \f{2}{3} \sum_j \Pi_j.
\ee

In terms of the Kasner exponents, this gives
\be \label{lqc-k}
k_i \to \t k_i = \f{2}{3} - k_i, \qquad k_\phi \to \t k_\phi = - k_\phi.
\ee
The sign change in $k_\phi$ simply signals the transition from a contracting to an expanding space-time.  This result can be compared to numerical simulations \cite{Gupt:2012vi}, which show agreement with the transition rule \eqref{lqc-k}.

\subsection{Bianchi II}
\label{ss.lqc-b2}

There is a quantization ambiguity in LQC that arises for Bianchi space-times with non-vanishing spatial curvature since there are inequivalent ways to represent the field strength operator in the Hilbert space.  In LQC, the preferred definition of a field strength operator is to express the field strength $F_{ab}$ in terms of a holonomy of the Ashtekar-Barbero connection around a loop of minimal area; this is known as the `F' loop quantization.  It is also possible to define a field strength operator from an operator corresponding to a difference of parallel transports of the Ashtekar-Barbero connection along a path parallel to $\oe^a_i$, or instead from an operator corresponding to a difference of parallel transports of the extrinsic curvature along a path parallel to $\oe^a_i$; these are called respectively the `A' and `K' loop quantizations.

For the Bianchi type II and type IX space-times, it is not known how to implement the heuristically preferred `F' loop quantization, but both the `K' and `A' loop quantizations are possible \cite{Ashtekar:2009um, WilsonEwing:2010rh, Singh:2013ava}.  It is not obvious which choice may better capture LQG effects, but the `K' loop quantization has a few nice properties.  First, the effective theory for the `K' loop quantization is better behaved than that of the `A' loop quantization, as the expansion and shear are always bounded for the `K' quantization but not for the `A' quantization \cite{Gupt:2011jh, Singh:2013ava}.  Second, for the closed FLRW space-time---where the `K', `A' and also the preferred `F' loop quantizations exist---the `K' loop quantization provides a better approximation to the `F' quantization than the `A' quantization \cite{Singh:2013ava, Corichi:2011pg}.  (In the flat FLRW and Bianchi I space-times, the `F', `K' and `A' loop quantizations all exist and are equivalent.)  Based on these advantages, I will consider the `K' loop quantization of the Bianchi type II and type IX space-times here, and leave for future work the study of the `A' loop quantization of these space-times.

For the `K' loop quantization, taking the lapse to be $N = V / \lo^3 = \sqrt{p_1 p_2 p_3} / \lo^3 = \exp[\sum_i \alpha_i]$ the LQC effective Hamiltonian constraint is \cite{Singh:2013ava}
\be
\mC_H = \mC_H^{(B.I)} + \lo^3 \: \f{e^{4 \alpha_1}}{32 \pi G},
\ee
where $\mC_H^{(B.I.)}$ is the LQC effective Hamiltonian constraint for Bianchi I \eqref{ch-b1}.  Note that the Bianchi II wall potential \eqref{U-b2} is not modified by any LQC effects.

Since the Bianchi II potential only becomes important for a short time during the transition between the two Kasner epochs, it is likely to be negligible during the LQC bounce, which also happens very rapidly.  Therefore, it is reasonable to assume the potential will be negligible during the LQC bounce for most choices of initial conditions, with the only exceptions being the highly fine-tuned cases where the Kasner transition happens simultaneously with the LQC bounce.

In the remainder, I will assume that the Kasner transition and LQC bounce occur at sufficiently different times so that it is possible to neglect the potential \eqref{U-b2} during the LQC bounce, and to neglect LQC effects during the Kasner transition generated by $U_{II}(\alpha_1)$.  For initial conditions giving dynamics where this is true, it follows that the transition rule for the Kasner exponents during the LQC bounce for the Bianchi II space-time is identical to the transition rule found for Bianchi I \eqref{lqc-k}.  In addition, as in this case LQC effects are negligible during the Kasner transition, the standard BKL transition rule \eqref{bkl-map} remains unchanged.

The classical dynamics for an expanding Bianchi II space-time, as reviewed in Sec.~\ref{ss.b2}, are relatively simple: at late times (and large volumes) the solution is nearly Kasner with $-1/3 < k_1 < 0$.  Evolving towards earlier times (and a smaller volume), the scale factor $a_1$ grows until the potential becomes important.  At this point, there is a transition between Kasner epochs following the BKL transition rule \eqref{bkl-map}, and then the system reaches the big-bang singularity staying in this second Kasner-like solution.

In the effective LQC theory, the singularity is resolved and so clearly the dynamics will be modified.  Again, the dynamics are described starting from a large volume solution, and evolving towards smaller volumes.  There are three possibilities:
\begin{itemize}
\item If the Kasner exponent $k_1$ is initially in the range $-1/3 < k_1^{(0)} < -2/7$ and a Kasner transition occurs before the LQC bounce, then $2/3 < k_1^{(1)} < 1$ after the Kasner transition, $-1/3 < k_1^{(2)} < 0$ after the LQC bounce, and there is no further Kasner transition after the bounce as $a_1$ is decreasing after the bounce.
\item If the Kasner exponent $k_1$ is initially in the range $-2/7 < k_1^{(0)} < 0$ and a Kasner transition occurs before the LQC bounce, then $0 < k_1^{(1)} < 2/3$ after the Kasner transition, $0 < k_1^{(2)} < 2/3$ after the LQC bounce, and since $a_1$ is increasing after the bounce, there will be one additional Kasner transition to $-1/3 < k_1^{(3)} < 0$ after the LQC bounce.
\item If there is no Kasner transition before the LQC bounce, then since the Kasner exponent is necessarily in the range $-1/3 < k_1^{(0)} < 0$, it follows that after the LQC bounce $2/3 < k_1^{(1)} < 1$.  Therefore, since $a_1$ will be increasing after the LQC bounce, there must be a Kasner transition after the bounce to $-1/3 < k_1^{(2)} < 0$.
\end{itemize}
Of course, at a Kasner transition or the LQC bounce, all three Kasner exponents change; here I focus on $k_1$ as it determines how many Kasner transitions there will be.

Note that it is impossible that there is no $U_{II}(\alpha_1)$-generated Kasner transition at all.  If there is no such Kasner transition before the LQC bounce, then since the post-bounce Kasner exponent $k_1$ will lie in the range $\tf{2}{3} < k_1 < 1$ the scale factor $a_1$ grows as $a_1 \sim |t-t_b|^{k_1}$ (with $t_b$ being the time the bounce occurs at), until the potential $U_{II}(\alpha_1)$ increases to the point where it causes a Kasner transition to occur.  So, there are three possibilities: a $U_{II}(\alpha_1)$-generated Kasner transition either before or after the LQC bounce, or two Kasner transitions, one before the LQC bounce and one after the LQC bounce.

\subsection{Bianchi IX}
\label{ss.lqc-b9}

The LQC effective Hamiltonian constraint for the Bianchi IX space-time, for the `K' loop quantization and $N = V / \lo^3$, is \cite{Singh:2013ava}
\be
\mC_H = \mC_H^{(B.I)} + U_{IX}(\alpha_i),
\ee
where $\mC_H^{(B.I.)}$ is the LQC effective Hamiltonian constraint for Bianchi I \eqref{ch-b1}, and $U_{IX}(\alpha_i)$ is the potential for the Bianchi IX space-time \eqref{U-b9}.

As for the Bianchi II space-time, in the limit that the potential is negligible during the LQC bounce and that LQC corrections are negligible during the potential-driven Kasner transitions, the LQC effective dynamics will have the same transition rule \eqref{lqc-k} during the LQC bounce, and will continue to follow the usual BKL oscillations \eqref{bkl-map} away from the bounce.

If $p_\phi \neq 0$, then the $U_{IX}(\alpha_i)$-generated Kasner transitions may stop before the bounce if all $k_i > 0$, but even in this case after the LQC bounce---during which the Kasner exponents change following \eqref{bkl-map}---at least two of the scale factors will be increasing, and so the potential will grow and cause more Kasner transitions.

For the vacuum case, each Kasner epoch can be parametrized by $(u, p_\Omega, v, \kappa)$ that, during $U_{IX}(\alpha_i)$-generated Kasner transitions, follow simple transformation rules relating subsequent Kasner solutions.  It is possible to derive the transformation rules relating these four parameters before and after the LQC bounce as well.

From \eqref{lqc-k}, it follows immediately that
\be \label{u-map-lqc}
u \to \t u = \f{\t k_{\rm max}}{\t k_{\rm mid}} = \f{\tf{2}{3} - k_{\rm min}}{\tf{2}{3} - k_{\rm mid}} = \f{u + 2}{u - 1},
\ee
where the relation \eqref{k-from-u} relating $u$ and the $k_i$ is used to obtain the last equality.  Note that $u$ transforms in the same manner no matter its value, unlike in the BKL map where it transforms differently depending on whether $u - 1 > 1$ or not.

Similarly, \eqref{pi-lqc-map} shows that
\be
p_\Omega \to \t p_\Omega = - p_\Omega.
\ee
The results follow directly from those obtained for the Bianchi I space-time with the only requirement that the Bianchi IX potential $U_{IX}(\alpha_i)$ remain negligible during the bounce.

More work is required to derive the transformation rules for $v$ and $\kappa$ since the $\alpha_i^{(0)}$ terms appear in their definition in addition to the Kasner exponents; this is what shall be done in the remainder of this part of the paper.

To do this, set $\tau=0$ at the last Kasner transition before the LQC bounce; the solutions for $\alpha_i(\tau)$ that are approximately valid from this Kasner transition until the LQC bounce are given by \eqref{alpha-min} and permutations thereof.  Since the Kasner transition (under the approximation used when deriving the transition rules for $v$ and $\kappa$) happens when the largest $\alpha_i=0$, it follows that one of the decreasing logarithmic scale factors must satisfy $\alpha_i(\tau=0) = 0$ (or equivalently, one of the two decreasing $a_i$ satisfies $a_i(\tau=0) = 1$), and so either $\alpha_{\rm mid}^{(0)}$ (if $v > 2$) or $\alpha_{\rm max}^{(0)}$ (if $1 < v < 2$) vanishes for the approximate Kasner solution before the LQC bounce.

Further, if $ v > 2$ and $\alpha_{\rm mid}^{(0)}=0$, then
\be
v = - \, \f{k_{\rm min} \, \alpha_{\rm max}^{(0)}}{k_{\rm max} \, \alpha_{\rm min}^{(0)}} + 2, \qquad
\kappa = \f{k_{\rm max}}{k_{\rm min}} \, \alpha_{\rm min}^{(0)},
\ee
implying that
\be \label{v2}
\alpha_{\rm min}^{(0)} = - \, \f{\kappa}{u+1}, \qquad \alpha_{\rm max}^{(0)} = - \kappa (v-2).
\ee
On the other hand, if $1 < v < 2$ and $\alpha_{\rm max}^{(0)}=0$, then
\be
v = - \, \f{k_{\rm mid} \, \alpha_{\rm min}^{(0)}}{k_{\rm min} \, \alpha_{\rm mid}^{(0)} - k_{\rm mid} \, \alpha_{\rm min}^{(0)}} + 1, \qquad
\kappa = k_{\rm max} \left( \f{\alpha_{\rm min}^{(0)}}{k_{\rm min}} - \f{\alpha_{\rm mid}^{(0)}}{k_{\rm mid}} \right),
\ee
in which case
\be \label{v1}
\alpha_{\rm min}^{(0)} = - \, \f{\kappa(v-1)}{u+1}, \qquad \alpha_{\rm mid}^{(0)} = - \, \f{(2-v) \kappa}{u}.
\ee

It is also helpful to note that at $\tau=0$, the relation \eqref{time} indicates that
\be
\ln \Big( 3 p_\Omega (t_{\tau=0} - t_o) \Big) = \sum_i \alpha_i^{(0)},
\ee
where $t_{\tau=0}$ denotes the instant $\tau=0$ in terms of proper time $t$.  Since $\sum_i \alpha_i^{(0)}$ is negative (as one of these constants has been assumed to vanish at the instant the Kasner transition occurs, while the other two are assumed to be negative so that their contribution to the potential \eqref{U-b9} is negligible), then $\ln 3 p_\Omega (t_{\tau=0} - t_o)$ is negative also.  In addition, as $t$ moves from $t_{\tau=0}$ to the bounce value $t_b$, the argument of the logarithm will shrink, and so the logarithm will become more negative:
\be \label{neg-log}
\ln \left( 3 p_\Omega (t - t_o) \right) \le \ln \left( 3 p_\Omega (t_{\tau=0} - t_o) \right) = \sum_i \alpha_i^{(0)} < 0,
\ee
for $t \in [t_{\tau=0}, t_b]$.  While this result is not essential for the calculations below, it shows which terms are positive and negative in the transformation maps derived below for $v$ and $\kappa$.

To make further progress, it is necessary to determine at what time $\tau$ the LQC bounce occurs.  This can be done under the assumption that the LQC bounce occurs when the space-time curvature reaches the Planck scale.  More specifically, in the effective theory it can be shown that the expansion $\theta = \d a_1/a_1 + \d a_2/a_2 + \d a_3/a_3$ is bounded \cite{Corichi:2009pp, Singh:2013ava}:
\be
\theta \le \theta_{\rm crit} = \f{3}{2 \gamma \sqrt\Delta},
\ee
with $\gamma$ and $\Delta \sim t_{\rm Pl}^2$ defined earlier following Eq.~\eqref{ch-b1}, and so here I will make the assumption that the LQC bounce occurs instantaneously when $\theta = 3/2 \gamma \sqrt\Delta$.

Under this assumption it follows from the solution \eqref{def-ai} for the $a_i$ that the bounce occurs at $t=t_b$ with
\be \label{def-tb}
|t_b - t_o| = \f{2 \gamma \sqrt\Delta}{3},
\ee
(with $t_b - t_o > 0$ if $p_\Omega > 0$ or $t_b - t_o < 0$ if $p_\Omega < 0$) and therefore, from \eqref{time},
\be
\tau_b = \f{1}{3 \, p_\Omega} \left( \ln 2 \, \gamma \sqrt\Delta \, |p_\Omega| - \sum_i \alpha_i^{(0)} \right),
\ee
recall from \eqref{neg-log} that $\ln 2 \, \gamma \sqrt\Delta \, |p_\Omega| < \sum_i \alpha_i^{(0)} < 0$.  From this, it follows that the value of the logarithmic scale factor at the bounce is
\be \label{alpha-b}
\alpha_i(\tau_b) = k_i \ln \left( 2 \, \gamma \sqrt\Delta \, |p_\Omega| \right) - k_i \sum_j \alpha_j^{(0)} + \alpha_i^{(0)}.
\ee

Of course there is some ambiguity here in choosing which quantity should be used to determine when the LQC bounce occurs (it would also be reasonable to consider the shear%
\footnote{The shear $\sigma^2 = [(H_1 - H_2)^2 + (H_2 - H_3)^2 + (H_3 - H_1)^2]/3$ is also bounded in the effective theory by $\sigma^2 \le \sigma_{\rm crit}^2 = 27 / (8 \gamma^2 \Delta)$ \cite{Corichi:2009pp, Singh:2013ava}.  If we were to assume that the LQC bounce occurs instantaneously when $\sigma^2 = \sigma_{\rm crit}^2$, this gives instead $|t_b - t_o| = 4 [1 - 16 \pi G p_\phi^2 / (36 p_\Omega^2 \lo^3) ] \gamma \sqrt\Delta / 9$.  Note that this value for $t_b$ is deeper in the Planck regime than \eqref{def-tb} (for all $p_\phi$, including the vacuum $p_\phi=0$ case); so the result \eqref{def-tb} follows also if one assumes that the LQC bounce occurs instantaneously when the first of $\theta = \theta_{\rm crit}$ and $\sigma^2 = \sigma_{\rm crit}^2$ occurs, since $\theta = \theta_{\rm crit}$ will always happen first.},
or a combination of the expansion and the shear) and also in the choice of the specific value at which the LQC bounce occurs (since it has not been shown that the bound is necessarily saturated by all solutions to the effective equations).  Nonetheless, the choice here is one strongly motivated by LQC, and has the further advantage of being relatively simple to implement.

It is now possible to calculate the transformation rules giving the values of $\t v$ and $\t\kappa$ after the LQC bounce from \eqref{v2}, \eqref{v1} and \eqref{alpha-b}.

Recall that $\t k_{\rm min} = \tf{2}{3} - k_{\rm max}$, $\t k_{\rm mid} = \tf{2}{3} - k_{\rm mid}$, and $\t k_{\rm max} = \tf{2}{3} - k_{\rm min}$, and therefore, under the assumption that the LQC bounce can be approximated by an instantaneous Kasner transition, $\t \alpha_{\rm max}(\tau_b) = \alpha_{\rm min}(\tau_b)$, $\t \alpha_{\rm mid}(\tau_b) = \alpha_{\rm mid}(\tau_b)$, and $\t \alpha_{\rm min}(\tau_b) = \alpha_{\rm max}(\tau_b)$, since only the labels change between the logarithmic scale factors during the LQC bounce.

Since the parameters $(u, p_\Omega, v, \kappa)$ are all (nearly) constant during a Kasner epoch, it is sufficient to calculate them at one time after the bounce; in this case it is clear that doing the calculations at the bounce time $\tau = \tau_b$ is the simplest choice.  Following from the arguments above, during the LQC bounce
\be
v \to \t v = - \, \f{1}{\t\kappa} \Big( \t\alpha_{\rm max} + (1 + \t u) \t\alpha_{\rm min} \Big) + 1
= - \, \f{1}{\t\kappa} \Big( \alpha_{\rm min}(\tau_b) + \f{2u+1}{u-1} \alpha_{\rm max}(\tau_b) \Big) + 1,
\ee
using \eqref{u-map-lqc} to replace $\t u$ by $(u+2)/(u-1)$, while $\t\kappa$ transforms as
\be
\kappa \to \t\kappa
= \t k_{\rm max} \left( \f{\t\alpha_{\rm min}}{\t k_{\rm min}} - \f{\t\alpha_{\rm mid}}{\t k_{\rm mid}} \right)
= - \, \f{2u + 1}{u - 1} \, \alpha_{\rm max}(\tau_b) - \f{u + 2}{u - 1} \, \alpha_{\rm mid}(\tau_b).
\ee
using the relation \eqref{k-from-u} to express the $\t k_i$ in terms of $\t u = (u+2)/(u-1)$.  The value of logarithmic scale factors at the bounce time, expressed in \eqref{alpha-b}, gives
\be \label{inter-v}
\t v = - \, \f{1}{\t \kappa (u - 1)} \left[ 2u \, \ln \left( 2 \, \gamma \sqrt\Delta \, |p_\Omega| \right) - (u+1) \, \alpha_{\rm min}^{(0)} - 2u \, \alpha_{\rm mid}^{(0)} + \alpha_{\rm max}^{(0)} \right] + 1,
\ee
and
\be \label{inter-k}
\t\kappa = - \, \f{2(u+1)}{u-1} \ln \left( 2 \, \gamma \sqrt\Delta \, |p_\Omega| \right) + \f{1}{u-1} \left[ 2(u+1) \, \alpha_{\rm min}^{(0)} + u \, \alpha_{\rm mid}^{(0)} + \alpha_{\rm max}^{(0)} \right].
\ee

Then, the transformation rules for $v$ and $\kappa$ can be determined from \eqref{v2} if $v > 2$, and from \eqref{v1} if $1 < v < 2$.  Interestingly, both cases give exactly the same transformation rules, namely
\be \label{lqc-kappa}
\kappa \to \t \kappa =
- \, \f{2(u+1)}{u-1} \ln \left( 2 \, \gamma \sqrt\Delta \, |p_\Omega| \right) - \f{\kappa v}{u-1},
\ee
and
\begin{align} \label{lqc-v}
v \to \t v &= - \, \f{1}{\t \kappa (u-1)} \left[ 2 u \, \ln \left( 2 \, \gamma \sqrt\Delta \, |p_\Omega| \right) + (3-v) \kappa \right] + 1 \nn \\ &
= \f{2 (2 u + 1) \, \ln \left( 2 \, \gamma \sqrt\Delta \, |p_\Omega| \right) + 3 \kappa}
{2 (u + 1) \, \ln \left( 2 \, \gamma \sqrt\Delta \, |p_\Omega| \right) + v \kappa}.
\end{align}
While these transformation rules are more complicated than those for $u$ and $p_\Omega$ due to the dependence of $v$ and $\kappa$ on the $\alpha_i$, they nonetheless have a relatively simple closed form.  Note that the Planck scale enters the transformation rules only for $v$ and $\kappa$: the transformation rules for $u$ and $p_\Omega$ are entirely independent of the Planck scale or the details of the quantum gravity effects during the bounce beyond the requirement that the equations of motion for the $\Pi_i$ all be the same during the bounce (in the limit that the potential $U_{IX}(\alpha_i)$ is negligible during the bounce).

It is also interesting that the same transformation rules apply for $(u, p_\Omega, v, \kappa)$ no matter their values, in contrast to the standard BKL map which is different depending on whether $u-1 > 1$ or not.  This may be due to the fact that for all LQC bounces, it is the labels $min$ and $max$ that are switched from the pre-bounce to the post-bounce era no matter the values of $(u, p_\Omega, v, \kappa)$, while in the standard BKL map the relabeling depends on whether $u-1 < 1$ or not.

Also, as can be seen from \eqref{neg-log}, the amplitude of the (negative) $\ln 2 \, \gamma \sqrt\Delta \, |p_\Omega|$ term is larger than the $v \kappa$ and $3 \kappa$ terms, so $\t v > 1$ and $\t \kappa > 0$, as required.  In some cases, it may be possible to entirely neglect the $\kappa$ terms with respect to the logarithmic term in which case the transformation rules will simplify further, but this may require special initial conditions due to the slow growth of the logarithmic function.

Another important point here is that the transformation laws for $u$ and $p_\Omega$ are more robust than those for $v$ and $\kappa$ because, as explained above, additional assumptions were required (namely, that one of the $\alpha_i^{(0)}$ vanishes at the last Kasner transition before the bounce, and that the LQC bounce occurs instantaneously when $\theta = \theta_{\rm crit}$) to derive the transformation maps \eqref{lqc-kappa} and \eqref{lqc-v}.

\subsection{Chaos?}
\label{ss.chaos}

An interesting question is whether the LQC effective dynamics for the vacuum Bianchi IX space-time are chaotic or not.  In LQC, the Bianchi IX space-time will undergo an infinite sequence of bounces and recollapses, with each bounce/recollapse cycle containing a finite number of Kasner transitions (during both the contracting and expanding phases), and, as a result, will have an infinite number of Kasner transitions when considering all bounce/recollapse cycles.

In the approximation that the full dynamics of the Bianchi IX space-time are well approximated by the sequence of Kasner parameters following the BKL and LQC transformation maps, and denoting the values of the $u$ and $v$ Kasner parameters%
\footnote{In general relativity, the $p_\Omega$ and $\kappa$ parameters do not contribute to the chaotic behaviour of the Bianchi IX space-time since they evolve monotonically (at least, in the contracting branch in the approach to the singularity that is most often considered) and there is no mixing in their trajectories.  While they are no longer monotonic in the LQC dynamics and so may generate some mixing from one cycle to another, this is likely to be subleading to the mixing in $u$ and $v$ that is responsible for the chaotic behaviour in general relativity.  For these reasons, I will ignore $p_\Omega$ and $\kappa$ here and focus on $u$ and $v$.}
during the $n^{\rm th}$ Kasner epoch by $(u_n, v_n)$, it is possible to calculate the topological entropy $H_T$, a coordinate-independent measure of the chaotic nature of a system, by counting the number $N(k)$ of periodic orbits after $k$ epoch changes, which is to say the number of pairs $(u_o, v_o)$ satisfying $(u_k, v_k) = (u_o, v_o)$, through the relation \cite{Bowen}
\be
H_T = \lim_{k \to \infty} \f{\ln N(k)}{k}.
\ee
By definition, $H_T$ can only be calculated in this context by allowing an infinite number of bounce/recollapse cycles of the Bianchi IX space-time, since there will only be a finite number of Kasner transitions during each cycle.  In other words, it is not possible to speak of chaos, at least in terms of $H_T$, if one only considers a single collapse-bounce-expansion cycle.  This is different from general relativity, where the topological entropy is non-vanishing already for the BKL transition map of a contracting Bianchi IX space-time as it approaches the cosmological singularity \cite{Barrow:1981sx, Chernoff:1983zz, Cornish:1996yg, Cornish:1996hx} without any need to consider any other phases of its evolution.

In this case, it will be difficult to explicitly calculate $H_T$ in this fashion for the infinite sequence of bounce/recollapse cycles of the Bianchi IX space-time in the LQC effective dynamics since the Kasner parameters, on their own, cannot predict whether the next Kasner transition should follow the standard maps \eqref{u-map} and \eqref{v-map}, or the LQC bounce map.  In addition, the BKL map is not a good approximation to the Bianchi IX dynamics around the recollapse, and so it will be necessary to go beyond the transition maps between the Kasner parameters and consider the full equations of motion near the recollapse point.  Following these considerations, it appears likely that numerical methods will be required to provide a definitive answer as to whether the Bianchi IX space-time is chaotic in LQC or not.

Nonetheless, the qualitative properties of the transformation map for the $u$ and $v$ variables at the LQC bounce do indicate that chaos is likely to be present in the effective LQC dynamics of the Bianchi IX space-time.  (Here I ignore the recollapse based on the assumption that if the truncation of the system to its near-bounce dynamics alone, repeated each cycle, are already sufficient to generate chaotic dynamics, then it seems reasonable to expect that adding the recollapse phases, if no attractor is present during the recollapse, will not remove the chaotic behaviour.)

To see this, consider the standard BKL map for $u$ and $v$ in the contracting Bianchi IX space-time:
\be \label{u-v-map}
(u_{n+1}, v_{n+1}) = \begin{cases}
(u_n - 1, v_n + 1) & {\rm if~} ~ u_n - 1 > 1, \\
((u_n - 1)^{-1}, 1 + v_n^{-1})  & {\rm otherwise}.
\end{cases}
\ee
During the contracting phase, take two solutions that are initially close with small differences in the Kasner parameters denoted by $\delta u$ and $\delta v$ (and defined to be positive).  As Kasner transitions occur, each time there is an era change $\delta u$ will increase and $\delta v$ will decrease.  This process, which also generates the mixing necessary for chaos, will continue until the LQC bounce.

After the bounce, the BKL map in the expanding space-time is the time-reverse of \eqref{u-v-map},
\be
(u_{n+1}, v_{n+1}) = \begin{cases}
(u_n + 1, v_n - 1) & {\rm if~} ~ v_n - 1 > 1, \\
(1 + u_n^{-1}, (v_n - 1)^{-1})  & {\rm otherwise}.
\end{cases}
\ee
Now the opposite happens: each time there is an era change, $\delta u$ decreases and $\delta v$ increases.  However, this will not be able to reverse the growth in the spread $(\delta u, \delta v)$ generated during the contracting phase, nor undo the mixing of the solutions, due to the physics at the bounce.

At the LQC bounce,
\be
(u_{n+1}, v_{n+1}) = 
\left( \f{u_n + 2}{u_n - 1}, \:
\f{2 (2 u_n + 1) \, \ln \left( 2 \, \gamma \sqrt\Delta \, |p_\Omega|_n \right) + 3 \kappa_n}
{2 (u_n + 1) \, \ln \left( 2 \, \gamma \sqrt\Delta \, |p_\Omega|_n \right) + v_n \kappa_n} \right).
\ee
First, a Taylor expansion shows that for $\delta u \ll 1 < u$ and away from $u=1$ where the expansion breaks down,
\be
\delta u \to \delta \tilde u = - \, \f{3}{(u-1)^2} \, \delta u.
\ee
As a result, $\delta u$ will grow for small $u$ and will be suppressed for large $u$.  In particular, as $u \to 1$, $\delta \tilde u$ grows unboundedly, while for large $u$, $\delta \tilde u \to 0$.  The cross-over point is at $u = 1 + \sqrt 3$, for which $\delta \tilde u = - \delta u$.

More importantly, at the bounce $v_{n+1}$ depends on $u_n$ and therefore the spread in $\delta u$ (which had been growing in the contracting phase) is now transferred, at least in part, to $\delta v$, precisely at the onset of the expanding phase when it is $\delta v$ that will now grow at each Kasner era change.  In exactly the same way, the mixing generated in $u$ during the contracting branch is now transferred to a mixing in $v$ at the LQC bounce, and after the LQC bounce the BKL map will generate further mixing in $v$.  As a result, the expanding branch cannot counteract or undo the spread or the mixing in the solutions generated by the contracting branch.  This is one reason why it appears very likely that the effective LQC dynamics for the Bianchi IX space-time are chaotic.

Another way to see this is by noting that the sensitivity to the initial conditions in \eqref{u-v-map}, which arises when $u_{n+1} = 1/(u_n-1)$ with $u_n - 1 < 1$, is not modified by quantum gravity effects.  Rather, quantum gravity effects add a new type of Kasner transition (which may or may not contribute to the chaotic behaviour itself) but it does not remove the source of the chaotic behaviour in the classical theory.  (That said, the bounce does ensure that there are only a finite number of Kasner transitions per bounce/recollapse cycle which makes it difficult to speak of chaos during one such cycle, but this is separate from the question of chaos for the infinite sequence of bounce/recollapse cycles).

It would be interesting to perform a numerical study of these dynamics, building on the earlier work \cite{Corichi:2015ala}, that can follow the effective dynamics through a number of bounce/recollapse cycles and extract the Kasner parameters at the times where a Kasner epoch is a good approximation to the dynamics, with the aim (i) to go beyond the transition maps between the Kasner parameters to the full effective dynamics in order to test the approximations used here and also to properly include the recollapse where the BKL map does not provide a good approximation to the full dynamics, (ii) to determine whether the dynamics are chaotic, perhaps by looking for fractal basin boundaries, and (iii) if the dynamics are found to be chaotic, then to extract estimates of, e.g., its topological entropy or fractal dimensions.

\section{Discussion}
\label{s.bkl}

In anisotropic space-times, in the effective framework the LQC bounce can be treated as a rapid transition of the Kasner exponents, following the simple law $k_i \to \t k_i = \tf{2}{3} - k_i$.  In Bianchi I space-times, this is the only transition of the Kasner exponents, while in Bianchi II space-times there is either one or two additional transitions that follow the well-known BKL transition map (at most one on either side of the bounce), and in Bianchi IX space-times, in each (of an infinite sequence of) bounce/recollapse cycles, there will typically be a number of BKL oscillations either side of the bounce, again following the BKL map.  These results were obtained assuming that the potential walls were negligible during the LQC bounce; further work is needed to determine what happens in the case when one of the potential walls is important during the LQC bounce.

Finally, due to the importance of the Bianchi models in the BKL conjecture, I end with a short discussion concerning the potential implications of these results for the BKL conjecture in the context of loop quantum gravity.

The BKL conjecture is that in the vicinity of a space-like singularity in general relativity, time-like derivatives dominate over space-like derivatives \cite{Belinski:1970ew}.  If correct, then in this limit the dynamics at each point decouple from each other and are given by ordinary differential equations; in particular, the dynamics at a generic space-time point will be those of a Bianchi IX model.  Thus, if the BKL conjecture is correct, and if the effective dynamics of LQC capture the key loop quantum gravity corrections to the Bianchi IX dynamics, then generic space-like singularities are resolved in loop quantum gravity.

However, a number of caveats are in order.  First, not all singularities in general relativity are space-like: under the assumption that the exterior region of the Kerr family is dynamically stable, there is a family of generic initial data that forms weak null singularities \cite{Dafermos:2017dbw}.  Second, even for space-like singularities, the BKL behaviour is only expected to appear asymptotically and there is no guarantee that BKL behaviour will appear before the Planck regime appears, which is when LQG effects are expected to be important; if in some space-times space-like derivatives are still comparable to time-like derivatives at the time when quantum gravity effects become large, then the BKL conjecture is not useful for understanding quantum gravity effects, at least in these particular space-times.  Third, even when the BKL behaviour is observed in numerical simulations of inhomogeneous space-times (and indeed, there is a significant amount of numerical evidence for the BKL conjecture in some space-times as reviewed in \cite{Berger:2014tev}), there typically do appear some `recurring spike surfaces', isolated non-generic surfaces where spatial derivatives cannot be neglected in comparison to time-like derivatives \cite{Rendall:2001nx, Andersson:2004wp, Lim:2007ta, Lim:2009dg, Heinzle:2012um, Coley:2016yuk}.  Clearly, the BKL conjecture fails in the immediate vicinity of these isolated surfaces.  Fourth, the LQC effective equations used in this paper are typically expected to be reliable when quantum fluctuations are negligible \cite{Rovelli:2013zaa}.  In quantum cosmology, this is often the case since the degrees of freedom in a homogeneous space-time correspond to large-scale observables (typically directional Hubble rates), for which quantum fluctuations are expected to be small so long as the physical scale factors always remain much larger than $\lp$ (in compact space-times, the scale factors are physical quantities that can be compared to $\lp$, and in non-compact space-times it is always possible to choose the fiducial cell so that this condition is satisfied).  However, in the context of the BKL conjecture, it is the dynamics at a point that are given by some Bianchi model, and quantum fluctuations in this case may become important in which case the effective equations may no longer be as reliable as they are for large homogeneous space-times.  Whether the effective equations can be trusted in such a context is a question that is left for future work.

With these caveats in mind, it is nonetheless interesting to consider what insight the results derived in the previous section can give with respect to the role of quantum gravity effects in at least some regions of high space-time curvature, if the LQC effective equations correctly capture the main LQG effects in a region of space-time close to where a space-like singularity would form in classical general relativity in the absence of quantum effects.

In the regions where the BKL behaviour occurs and the loop quantum gravity effects at each point are captured by the LQC effective equations, the classical space-like singularity is resolved point by point, and at each point the energy density, the expansion and the shear are all bounded \cite{Corichi:2009pp, Singh:2013ava}.  Furthermore, the dynamics at each point is given by a sequence of Kasner epochs linked by Bianchi II transitions, and the singularity is replaced by a non-singular bounce governed by the LQC effective equations, which can be approximated by a new type of Kasner transition, as explained in the previous section.

In such space-time regions, it is interesting to consider the dynamics of two nearby points.  Assuming that the metric is continuous (or if the metric is an emergent quantity from the fundamental quantum theory, then assuming that the emergent metric is continuous at length-scales where it is possible to speak of a metric), if these two points are sufficiently near each other then their Kasner parameters $(u, p_\Omega, v, \kappa)$ will, at least initially, be close to each other.  Will the Kasner parameters at these two points remain close to each other, or will they be driven far apart by the LQC-corrected BKL dynamics?  In other words, will inhomogeneities grow uncontrollably?  This issue is clearly closely related to the question of chaos in the Bianchi IX dynamics.

Based on the discussion in Sec.~\ref{ss.chaos}, the dynamics in such a situation will be sensitive to the initial conditions at Kasner era changes and potentially at the LQC bounce.  Therefore, the question becomes how many Kasner era changes should be expected between the onset of the BKL behaviour and the LQC bounce; the more Kasner era changes there are, the more the dynamics will drive apart the solutions for the gravitational fields at neighbouring points.  Also, the value of $u$ at the bounce may play a role, since for $u > 1+ \sqrt{3}$ the LQC bounce will suppress inhomogeneities in $u$, but will amplify inhomogeneities in $u$ if $u < 1 + \sqrt{3}$.

Numerical studies indicate that there are typically a small number of Kasner era changes between the onset of the BKL behaviour and the point when the Planck scale is reached (see, e.g., \cite{Weaver:1997bv}, and also \cite{Doroshkevich} for analytical results in the homogeneous case of the Bianchi IX space-time, although focused on length scales rather than curvature scales).  If there are indeed only a small number of Kasner era changes, then it will be reasonable to expect that although the solutions at neighbouring points will separate and inhomogeneities will grow, they will not do so uncontrollably---instead the amplitude of these effects will be limited.  However, it may well be necessary to check how many Kasner era changes there are on a case by case basis in order to quantify how strongly the solutions at nearby points separate during the bounce.

Of course, this discussion remains qualitative and more work is needed to understand how to perform a full LQC/LQG quantization of inhomogeneous space-times, perhaps by using the Hamiltonian formulation of the BKL conjecture in connection variables \cite{Ashtekar:2008jb, Ashtekar:2011ck}, in order to determine the validity of these arguments and to provide quantitative answers to the question of the fate of space-like singularities in loop quantum gravity.

\newpage

\acknowledgments

I thank Masooma Ali, Viqar Husain and Sanjeev Seahra for helpful discussions and comments on an earlier draft of the paper.

\raggedright

\end{document}